# Electrochemical and XAS analysis of the ruthenium-dioxide material, a catalyst for Dimensionally Stabilized Anodes

Sujit K Mondal[a≠]

[a]*Past affiliation-Department of Chemistry and Chemical Biology, Northeastern University Boston Massachusetts 02115 USA.*

[≠]*Present affiliation- Biomaterial and Biomimetic Centre, University of Texas, 6516 MD Anderson Blvd, Texas-77030, USA.*

Corresponding authors. Tel: +1713 500 4381

*E-mail address:* Sujit.K.Mondal@uth.tmc.edu *(S. K. Mondal). sujitharvard@gmail.com*

**Abstract**

This manuscript represents an investigation of structure, property and electrochemical activity for commercially available $RuO_2$ coated titanium electrode, known as DSA. The investigation mainly aimed at XAS studies to understand the hidden structural information for $RuO_2$. Two different (150 micron) and a (6 micron) titanium metal coated with $RuO_2$ was exposed at ex-situ and in-situ XAS experiment to find out its electrocatalytic property and corrosion during chlorine generation reaction. A detailed survey of XANES and EXAFS showed that the stable $RuO_2$ formed a nano cluster during chlorine generation reaction and these clusters produced a distorted rutile structure with an enhanced electrocatalytic activity. These nano clusters possibly the cause of generating surface heterogeneity of the present ruthenium dioxide material (active catalyst for DSA). With the help of electrochemical impedance spectroscopy we are able to detect the surface heterogeneity, this surface heterogeneity is having key role for outstanding electrochemical performance of these DSA electrodes.

## Introduction

Four decades have elapsed since the instant when the first titanium anodes with a metal oxide active coating, known as dimensionally stable anodes (DSA), were created. Among these anodes, the electrodes with coatings based on mixed ruthenium and titanium oxides (ORTA) and the latter with additives of doping oxides of non-noble metals with partial or complete replacement of ruthenium oxides of other noble metals in their coatings enjoyed extensive practical application. By now these anodes practically and completely replaced the graphite anodes that had been used in the chlorine industry. This was due to the fact that they substantially exceed the latter by their catalytic activity, selectively, and corrosion resistance. The presence of such characteristics of anodes explained the tremendous interest that was exhibited in studying the corrosion and electrochemical behavior of anode of this type. The DSA anodes are very important in the context chlorine industry and hydrogen-halogen based reversible fuel cell. Commercial chloride oxidation electrodes are based on DeNora's Dimensionally Stable Anode (DSA), which is an oxide of the $Ru_xTi_{1-x}$ alloy, with x typically greater than 30% [1]. DSAs have superb electrocatalytic properties for chloride oxidation at highly anodic potentials. The DSA electrode kinetics of the chlorine oxidation reaction have been studied extensively [2-5]. Hydrogen-halogen based regenerative fuel cells from the 1980s employed comparable electrode materials to DSAs [7]. Due to immense importance of DSA As a result of performed investigations, kinetics was studied and the mechanism of the evolution of chlorine and oxygen was investigated.

From the literature survey we found that scientists have used $RuO_2$ rutile phase oxide material for both chlorine and oxygen generation reaction. Scientists have already tried other oxide material to replace the $RuO_2$ but they found that the activity (performance) was one the main issue related to the oxide material [8]. The different values of tafel slope was reported during chlorine generation reaction by varying the calcinations temperature and also by changing the concentration of chloride containing electrolyte [8,9]. At this context $RuO_2$ become more interesting material as it shows some high electrocatalaytic property and high stability during chlorine generation reaction. Scientists have also used mixed oxide [8] for DSA but the best results obtained for $RuO_2$ and that's the main reason scientists are interested to explore DSA by using the $RuO_2$ catalyst.

It was also found that very few reports were available for electrochemical impedance spectroscopy, which was not directly related to catalyst $RuO_2$ coated over titanium, here in this manuscript a logical conclusion was derived from electrochemical impedance study of $RuO_2$ during chlorine oxidation reaction. The main active material for DSA was $RuO_x$ (nominally $RuO_2$), coating solution was supplied by DeNora tech company for our present study. It was routinely prepared by thermal decomposition of $RuCl_3$ $nH_2O$ (with n= 2, 3) so that chlorine remained in the oxide during the preparation route. The chlorine content decreases as the time of calcinations increases. Under practical conditions, its concentration may be an order of a few weight percent [9].

Reports about EIS technique for the $RuO_2$ material was found on rotating ring disc electrode [10] and $RuO_2$ nano needles on glassy carbon electrode [11], in the present study we used the direct catalyst material coated over titanium electrode to study electrochemical impedance spectroscopy to evaluate the double layer capacitance and charge transfer resistance. These parameters were obtained by fitting only semicircle of EIS spectra with the software available from Autolab Instrument. We were able to give new information (origin of new degrees of geometry) about the oxide material, mainly $RuO_2$ coated electrode. XAS experiment showed some interesting phenomenon, formation of $RuO_2$ nano clusters of both dimers and trimers on the Ti surface during chlorine generation reaction, which possible a key factor for enhancement of catalytic properties for $RuO_2$. In this manuscript we are trying to co-relate the structure property relationship for $RuO_2$ during chlorine generation reaction and trying to answer the following questions, which are still not being answered properly to the best of author's knowledge.

1) Can it be possible to co-relate the structure property relationship from In-situ and Ex-situ XAS studies for $RuO_2$ during chlorine generation reaction?
2) Is it really the rutile phase of $RuO_2$ responsible for its higher electrochemical activity?

A detailed analysis and understanding of the above mentioned questions were explained in the following manuscript.

**Experimental**

Electrodes for this study were provided curtesy of De Nora Tech (a subsidiary of Industries De nora S.p.A.). The dimension of the titanium was (1.0 X 1.0) cm$^2$. To follow up chlorine oxidation reaction in sequence the experiments were carried out in two different titanium current collectors coated with $RuO_2$. The maximum sample loading level obtained for 150 micron was 12-15g m$^{-2}$ and 6 micron was 12-15g m$^{-2}$ for XAS experiment.

All the electrodes were used for cyclic voltammetry experiments. A three electrode glass cell was assembled and used for cyclic voltammetry experiment. Argon gas was purged for 20 minute to the electrolyte solution to make it an oxygen free medium. Working potential window for ruthenium di-oxide was experimentally evaluated depending upon the hydrogen and oxygen generation reactions. The ruthenium di-oxide showed its characteristic voltammograms with a high double layer charging. The cyclic voltammetry experiments were performed for $RuO_2$ electrode in 0.5(M) $Na_2SO_4$ and 0.07(M) NaCl solution with Ag/AgCl as the reference electrode. The two different sample specimens for $RuO_2$ (over 150 micron and 6 micron titanium) were prepared for SEM experiment. Due to the morphological features, the $RuO_2$ electrode was referred as 'mud cracked morphology'.

Electrochemical impedance spectroscopy experiment was performed with the electrodes. Three electrode configuration cell was set up for EIS experiments. The AC impedance experiment was carried out with a wide frequency range starting from 100 kHz to 10 mHz range. The EIS experiments for $RuO_2$ electrode were performed by conditioning the electrode at 0.8 V (before chlorine generation potential) to a high potential of 1.4 V (beyond chlorine generation potential) with respect to Ag/AgCl electrode. The Nyquist diagram was obtained for this electrode with a wide frequency range starting from 100 kHz to 10 mHz. Due to vigorous chlorine evolution at 1.4 V the spectrum was having little scattered region at low frequency zone, possibly due to chlorine gas generation reaction. All the EIS spectra were fitted with the Autolab software. The electrode potential was read against Ag/AgCl as reference electrode for $Na_2SO_4$ + NaCl mixed solution for $RuO_2$ electrode.

The XAS ex-situ experiments were performed for the $RuO_2$ sample coated over 150 micron titanium. X-ray absorption spectroscopy data on the samples were collected at the National Synchrotron Light Source, Brookhaven National Lab, Upton NY at beam line X-11A. All data were collected at the Ru K-edge in transmission mode. The data consisted of spectra for two samples, one cycled to a potential *below* that at which no chlorine evolution, (sample II) and the other, to a potential *above* the chlorine evolution potential (sample I). Both spectra were collected *ex-situ* in air, at room temperature (22 °C). The In-situ experiments were performed over 6 micron titanium substrate coated with $RuO_2$ at various potential ranges starting with 0.65V, below the chlorine evolution and above the chlorine generation potential eg. 1.5V, all the potential values were measured with respect to Ag/AgCl. Data reduction of the EXAFS spectra was performed, pre-edge and postedge backgrounds were subtracted from the XAS spectra and the results were normalized with respect to edge height. Curve fitting was performed with Artemis and IFEFFIT software using ab initio calculated phases and amplitudes using particular the program. These ab initio phases and amplitudes were used in the EXAFS equation:

$$\chi(k) = S_0^2 \sum_j \frac{N_j}{kR_j^2} f_{eff_j}(\pi, k, R_j) e^{-2\sigma_j^2 k^2} e^{-2R_j/\lambda_j(k)} \sin(2kR_j + \phi_{ij}(k)) \quad (1)$$

$f_{eff_j}(\pi, k, R_j)$      1(a)

$e^{-2\sigma_j^2 k^2}$      1(b)

$e^{-2R_j/\lambda_j(k)}$      1(c)

$\sin(2kR_j + \phi_{ij}(k)$      1(d)

The neighboring atoms to the central atom(s) are divided into *j* shells, with all atoms with the same atomic number and distance from the central atom grouped into a single shell. Within each shell, the coordination number $N_j$ denotes the number of neighboring atoms in shell *j* at a distance of $R_j$ from the central atom. Equation 1(a) is the *ab initio* amplitude function for shell *j*, and the equation 1(b) represents Debye-Waller term accounts for damping due to static and thermal disorder in absorber-back scatterer distances. Equation 1(c) represents mean free path term, reflects losses due to inelastic scattering, where $\lambda_j(k)$ is the electron mean free path. The oscillations in the EXAFS spectrum are reflected in the sinusoidal term equation 1(d), where 2nd part of the sine function is the ab initio phase function for shell *j*. $S_0^2$ is an amplitude reduction factor due to shakeup/shake-off processes at the central atom(s). The EXAFS equation was used to fit the experimental data using N, R, and the EXAFS Debye-Waller factor ($\sigma^2$) as variable parameters. EXAFS curve fitting procedures and the estimations of the uncertainty in the parameters from the fits are described in detail in the XAS section of results discussion.

## Results and Discussion

1) **Dynamic method**

a) **Cyclic Voltammograms for $RuO_2$ electrode:** Cyclic voltammograms for $RuO_2$ were very interesting, at first we observed oxygen generation (1.1V) followed by chlorine evolution potential around 1.2V and also hydrogen evolution at almost -0.05 V versus Ag/AgCl electrode, it was shown in Fig.1. The voltammogram have a potential window for redox charging, 0.0 V to 1.0 V with respect to Ag/AgCl reference electrode also a faradaic reaction zone beyond 1.15V, this is predominantly for chloride oxidation reaction. The cyclic voltammetry experiments were performed with various scan rates in 0.0 to 1.0 V potential zone, as shown in Fig.2, a typical capacitor type voltammogram was obtained which was very similar to $RuO_2$ with mud-cracked morphology. The voltammograms were almost symmetrical in nature, this was specifically due to redox charging/sodium ion diffusion into cracked sites and interstitial sites of $RuO_2$. This phenomenon was observed for all the 2 substrates of titanium coated with $RuO_2$ irrespective of the thickness. The maximum anodic current value 0.5 mA per cm$^2$ (geometric area) was obtained for 10 mVs$^{-1}$ scan rate for $RuO_2$ in coated over 150 micron titanium. . A Linear relationship between anodic charge and scan rate, explained diffusion controlled reaction for oxide catalyst. The diffusion of sodium ion (as sodium ion present in the electrolyte) was probable case inside the oxy-hydroxy layer of the catalyst [12,13]. Possibly, there could be two mechanisms for the anodic redox charging reaction under one electron transfer reaction scheme.

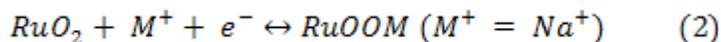
$$RuO_2 + M^+ + e^- \leftrightarrow RuOOM \ (M^+ = Na^+) \quad (2)$$

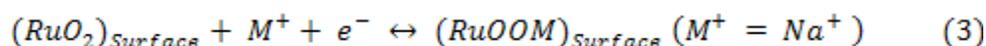
$$(RuO_2)_{Surface} + M^+ + e^- \leftrightarrow (RuOOM)_{Surface} \ (M^+ = Na^+) \quad (3)$$

From the literature we found that diffusion limited redox charging is possible for crystalline oxide material and amorphous oxide material [12,13]. The redox charging for amorphous oxide followed an adsorption of M$^+$ ion at the surface sites of the oxide material; this process described in equation (3). The redox charging for crystalline ruthenium-dioxide followed a probable cationic diffusion inside cracks and crevices of the oxide material; was described in equation (2)[10,12,13]. Our DeNora Tech prepared ruthenium dioxide had the rutile phase (confirmed by XAS-experiment also), we believed that redox charging followed the mechanistic pathway of equation (2). High anodic charge and capacitance were observed for ruthenium-dioxide, which in overall explained the fact that catalyst was having smaller particle size and high surface area (higher electrochemical active sites). The ion adsorption became easier due to high surface area of the active oxide catalyst, mainly chloride ion during anodic oxidation process of chloride to chlorine gas generation. Fig.3 is the scanning electron micrograph of ruthenium dioxide, it has well known mud cracked morphology. The cracks were manifestation of macro roughness, while pores/crvice constituted micro roughness. This type of morphology enhanced the diffusion process of sodium ion (easier cationic diffusion into defect sites/interstitial sites, mainly macro roughness and micro roughness)[10] and a high value of double layer charging current was observed during cyclic voltammetry experiment. .

## 4) AC Technique

**Electrochemical Impedance Spectroscopy:** Electrochemical impedance spectroscopy is an useful tool to investigate the charge transfer process (faradic reaction) at electrode-electrolyte interface, double layer capacitance, diffusion controlled electrode kinetics/psedocapacitive behavior. A model Nyquist plot of EIS showed in Figure 4(a), all these processes at different frequency interval were explained in this figure. Figure 4(b) is the model equivalent circuit was used to fit the semi circle of the nyquist spectrum. The classical approach to impedance modeling, based on the assumption that the interface model should contain lumped elements only, such as R, C, L and Warburg impedance, cannot be applied to our system. This type of modeling suits homogeneous systems but with system, having high surface heterogeneity needs constant phase element to model the impedance spectra. The constant phase element (CPE) which is distributed by its very nature, is such an element, it sums the impedance response of the distributed process in a single expression. It has peculiar property of causing a frequency impedance phase shift between an applied alternating potential and current response. The impedance of CPE described in equation 4.

$$Z_{CPE} = A^{-1} (j\omega)^{-n} \qquad (4)$$

In equation 4, A represents a proportionality factor and n has a meaning of phase shift. Lately, many authors use CPE for modeling of frequency dispersion behavior corresponding to different physical phenomena such as surface heterogeneity, which results from surface roughness, impurities, dislocations, grain boundaries fractality [15-17] etc. In this article n should be considered as surface heterogeneity, which is a measurement of fractality. A fractal is an object with complex structure, revealing new details at increasing degrees of magnification. Fractal dimension of an electrode can be measured by using degree of depression of semicircle in Nyquist spectra, by following expression written in equation 5.

$$n = \frac{1}{D_{fs}-1} \qquad (5)$$

$D_{fs}$ can be measured as fractal dimension of the surface and can take value between 2 to 3. $D_{fs}$ represents smooth surface with a values close to 2 and rough surface with a value close to 3. Figure 5(a) was impedance spectrum of $RuO_2$ collected at 0.8V and figure 5(b) impedance spectrum collected at 1.4V. The double layer capacitance and charge transfer resistance were calculated by fitting only the semi circle of the Nyquist spectra in AUTO lab software using constant phase element.

In Nyquist plot, the charge transfer process for electrode-electrolyte surface was occurred at higher frequency region, represented by a semi circle. A straight line at lower frequency region with an angle around $45^0$ to the real axis represented diffusion controlled kinetics, known as the Warburg component, and the sharp increase of the spectrum after the Warburg component was due to pseudo capacitive behavior of the oxide, mainly due to diffusion of ions. This diffusion layer thickness can vary in size from finite to semi-infinite length. The thickness of diffusion layer increased gradually as we increase the anodic potential. As diffusion layer thickness became bigger warburg impedance part touched the real axis at 1.4V, $2^{nd}$ semicircle in nyquist spectrum of figure 5(b) was a consequence of the increase thickness of diffusion layer. Figure 6(a) showed a behavior of charge transfer resistance with anodic potential, charge transfer resistance for overall electrochemical reaction had decrease in trend with an increase in applied potential. This behavior possibly explained the phenomenon of facile chloride ions adsorption of over oxides surface at higher potential. Facile adsorption occured when higher numbers of adsorption sites were available. Higher numbers of adsorption sites were generated due to the enhancement of surface heterogeneity/fractal dimension ($D_{fs}$). Higher number of adsorption sites over oxides surface opened up more room for chloride ions. When chloride ion concentration increased at the electrode–electrolyte due to enhanced surface heterogeneity; thickness of total double layer capacitance would also alerted. Figure 6(b) represented an overall increase in trend of double layer capacitance with anodic potential. Enhancement of double layer capacitance occurred due to enhancement of overall electrochemical surface area of the oxide material, as surface heterogeneity was also increased.

The figure 7(a) and 7(b) represented the change in surface heterogeneity with the change in anodic potential. The different n values (obtained from fitting with CPE-equivalent circuit) were plotted with anodic potential in figure 7(a), enhancement of fractal dimension with respect to anodic potential and n value obtained from fitting nyquist spectra was plotted in together in figure 7(b). From the plot it was clear that at given n (CPE) of 0.5, $D_{fs}$ reached a value of close to 3, a very high surface heterogeneity was obtained. Figure 7(b), it was clear that high anodic potential generated an enhancement of the surface heterogeneity which resulting a facile adsorption of chloride ions and decrease in the overall charge transfer resistance for the electrochemical process.

**5) Ex-situ XAS Analysis:**

**The XANES region:** All data were processed using the IFEFFIT suite [18-20] (v.1.2.9). The scans were first normalized between 0 and 1 for XANES analysis and the reference scans of the samples were calibrated and aligned to the standard Ruthenium K-edge (22117 eV)[21]. The shifts thus observed represent the actual shift in the edge energy due to the oxidation state of Ruthenium in the samples. Fig.8 showed the XANES region for the two samples and a standard Ruthenium foil, which was expected to be in its elemental state, viz. Ru(0). The edge energy for samples I and II were located at 22123.72 eV and 22125.87 eV. The most apparent difference between the two samples is the difference in the white-line intensity of the main peak centered around 22140 eV. The higher intensity in the case of sample I indicates that this sample is oxidized to a larger extent than sample II. It correlated well with the fact that sample I was cycled to a higher potential than sample II and thus suggests a clear, local morphology change between the two samples. These changes are discussed in greater detail in the section that follows.

| Sample 1 | Ru-O | Ru-Ru (1$^{st}$ Shell) | Ru-Ru (2$^{nd}$ Shell) |
|---|---|---|---|
| N | 7.08 | 0.87 | 3.19 |
| $\Delta E$ (eV) | -32.12 | 25.76 | 16.00 |
| r (Å) | 1.955 | 2.564 | 3.398 |
| $\sigma^2$ (Å$^2$) | 0.0041 | 0.0043 | 0.004 |

| Sample 2 | Ru-O | Ru-Ru (1$^{st}$ Shell) | Ru-Ru (2$^{nd}$ Shell) |
|---|---|---|---|
| N | 5.21 | 0.57 | 2.22 |
| $\Delta E$ (eV) | -28.97 | 31.31 | 17.61 |
| r (Å) | 1.961 | 2.596 | 3.399 |
| $\sigma^2$ (Å$^2$) | 0.0041 | 0.0043 | 0.004 |

Table 1. Ru EXAFS First and second shell fits on both samples I and II using Ru-O and Ru-Ru paths

**EXAFS Analysis:** The data were processed to obtain the oscillatory chi(k) function by removing the background above the edge and was fit using the standard procedure used in EXAFS analysis. Briefly, the energy in electron volts (eV) was converted to k-space over the region from 2 to 13.25 Å$^{-1}$. The data were then k$^2$-weighted and Fourier-transformed to produce a partial radial distribution (RDF) function around Ruthenium. The fits were carried out in the region 1.193 Å < r < 3.633 Å using a Kaiser-Bessel window spanning this data range in r-space. The partial radial RDFs of the samples were analyzed using the FEFF code (v. 6.0) incorporated in the program ARTEMIS [22] to obtain average bond distance (r), coordination number (N) and the Debye-Waller factor ($\sigma^2$) for the nearest Ru-O and Ru-Ru neighbors. In all, three paths (1 Ru-O and 2 Ru-Ru) were fit to the samples using the four standard parameters for each path viz., the coordination number (N), energy ($\Delta E$), radius of path ($\Delta r$) and the Debye-Waller factor ($\sigma^2$), which was a measure of thermal disorder in the samples. All fits were carried out in r-space and the fits for samples I and II were shown in fig. 11 and 12 respectively. The comparison of pseudo radial distribution function and oscillatory chi(k) function for both samples with Ru foil were described in Fig. 9 and Fig. 10. The results of the fits were shown in Table 1.

**Discussion of results:** From the XANES region (Fig.8), it is seen that the Ruthenium in sample I is oxidized to a larger extent than that in sample II. This is seen in both, the edge energy shift ($\Delta E \sim 2$ eV) between the two samples and in the white line intensity. Although the overall shift in energy from that of Ru$^0$ is close to that found for Ru$^{4+}$ (22.126 keV), McKeown et.al.[24] show that it is difficult to distinguish between Ru$^{3+}$ and Ru$^{4+}$ in ruthenium oxides from just the XANES region. It is thus possible that there is a certain fraction of both oxidation states in the samples.

The data was initially fit using an anhydrous, crystalline $RuO_2$ ($Ru^{4+}$) cluster wherein each $Ru^{4+}$ ion in the $RuO_6$ octahedra was bonded to four oxygen ions at a distance of 1.984 Å and two axial oxygen ions at a distance of 1.942 Å. However, the fits were of moderate quality. This could be due to one of the following reasons –

  i.    The samples I and II do not possess a perfectly rutile-like $RuO_2$ environment.
  ii.   The data quality and range over which the analysis was carried out limits the bond distance resolution to ca. 0.2 Å and thus, the fits with the two oxygen distances separated by under 0.05 Å does not yield good fits with acceptable parameters.

From observing the XANES region of the spectra and comparing it with those available in the literature [23-28], it is more likely that the samples are unlike standard $RuO_2$. The results from the fits indicate that the $Ru(n^+)$ ions are present in an octahedral environment and surrounded by around 6 oxygen ions at a distance of ca. 1.95-1.96 Å. The bond distance obtained is intermediate to the Ru-O distances in $RuO_2$. The Ru-O coordination increases for the more highly oxidized sample I and is consistent with XANES results. The nearest significant Ru-Ru coordination seems to be at ca. 3.4 Å, which is considerably larger than the closest Ru-Ru distance in $RuO_2$ at 3.107 Å. This peak representing Ru-Ru scattering is quite broad and suggests a distribution of Ru-Ru nearest neighbor distances.

The fit between 2-3 Å is in moderate quality and could be due to the interaction of Ru with the Ti substrate. As a first attempt in understanding this catalyst structure, including too many paths was avoided to minimize complexity in the analysis and interpretation of fit results.

**6) In-situ XAS Analysis:** The in situ data discussed in this section were collected at the Ru K-edge in fluorescence mode using a standard PIPS detector and sample used was $RuO_2$ coated over 6 micron titanium foil.

XANES analysis –

Fig.13 XANES region of spectra collected at various potentials including on set before and after chronoamperometric (CA) studies (data at 1.2V). The in-situ data collected in the XANES region was interesting in that there was virtually no change in the spectra across the various potentials. The data indicates that the material is in the same oxidation state, i.e. highly oxidized state at all potentials was investigated. This was not surprising given that the material being investigated was primarily $RuO_2$ and was thus expected to be fully coordinated to oxygen and likely

independent of the applied potential. An edge shift of ca. 8.6 eV from that of the foil ($Ru^0$) was observed.

EXAFS analysis –

Fig.14 and Fig. 15

The raw data in k-space are shown in Fig. 15. The data were analyzed between $2 < k < 13.95$ Å$^{-1}$ and $1 < r < 3.579$ Å. A k-weight of 2 was used for all the fits. The pseudo-radial distribution function showing the data transformed into r-space are shown in Fig.14. The large peak centered on 1.6 Å was due to backscattering from the oxygen atoms present in the first shell of the $RuO_2$ catalyst particles. There seems to be little change in the Ru-O coordination with increase in potential between 0.65V and 1.2V. Further, the coordination did not seem to change noticeably even after chronoamperometric studies at 1.2 V for 15min, suggesting that the material is quite stable to electrochemical corrosion at high potentials. This peak however did increase in intensity slightly at 1.5V due to increased oxidation of the material.

As seen previously in the ex-situ data, the region between 2-3 Å is somewhat complicated and therefore, unclear. The peak at 2.2 Å is most probably due to a longer Ru-OH bond distance while the peak between 2.6-2.8 Å appears to be due to first shell Ru-Ru backscattering as it decreased slightly with potential, possibly due to increased Ru-O coordination or (Ru-Cl). A very interesting feature of the spectra at 1.2 V (after CA study) and 1.5V when compared with the other spectra at lower potentials is the appearance of a small peak at ca. 2.0 Å. This peak is suspected to be due to Ru-Cl scattering and is therefore only expected at higher potentials and after significant evolution of chlorine gas from chloride ions, whereby Cl$^-$ species participated in diffusion through the material causing Ru-Cl scattering. In the data at 1.5V, it was only a small shoulder whereas at 1.2V after the CA study, it was a small, though distinct peak because at 1.2 V continuous chlorine generations reaction was happened for 15 minutes.

The peak around 3.1 Å is most definitely due to Ru-Ru scattering at a distance of 3.58 Å as seen in anhydrous $Ru(OH)_2$. There is no change with potential indicating that any changes with potential are chiefly occurring within the first shell (below 3 Å) of the catalyst material. The spectrum at 0.65 V. and 1.5 V has been fitted with the Artemis software and results were tabulated in the following table. Fitted spectrum obtained from 0.65 V was described in Fig.16. It was clear from the table that there was a considerable amount of increase in co-ordination number from 0.65 V to 1.5 V which reflected the formation of (Ru-O)$_n$ nano cluster and thus enhancement of the catalytic activity for RuO$_2$ catalyst coated over titanium metal.

| Sample (0.65V) | Ru-O | Ru-Ru (1$^{st}$ Shell) | Ru-Ru (2$^{nd}$ shell) |
| --- | --- | --- | --- |
| N | 1.64 | 0.68 | 0.6 |
| ΔE(eV) | 10.37 | -10.35 | 10.2 |
| r(Å) | 1.91 | 3.11 | 3.54 |
| σ$^2$ (Å$^2$) | 0.008 | 0.0085 | 0.004 |

| Sample (1.5V) | Ru-O | Ru-Ru (1$^{st}$ Shell) | Ru-Ru (2$^{nd}$ shell) |
| --- | --- | --- | --- |
| N | 7.72 | 1.335 | 1.2 |
| ΔE(eV) | 10.3 | -16.47 | -10.2 |
| r(Å) | 1.8 | 3.019 | 3.53 |
| σ$^2$ (Å$^2$) | 0.006 | 0.0069 | 0.002 |

**Conclusions:-**

1) The results reported in this manuscript clearly suggested that the electrochemical activity of RuO$_2$ originated from its specific crystalline rutile nature and also from surface heterogeneity. The cyclic voltammograms showed high pseudo capacitance due to redox charging with the possible mechanism as follows (RuO$_2$ + M$^+$ + e$^-$ ↔ RuOOM ). M$^+$ = Na$^+$, H$_3$O$^+$ etc.

2) Impedance spectra depicted a change in double layer capacitance for RuO$_2$ material, which we believed an enhancement of electro active surface area of the oxide material due to introduction of more surface heterogeneity. By modeling the EIS spectra with constant phase element, we were able to predict the surface heterogeneity of the oxide very clearly.

**3)** From ex-situ XAS study we found that the primary difference between the two samples I and II, sample I was oxidized to a larger extent than sample II. This information was clearly visible from both, the XANES region and the EXAFS analysis and was consistent with the experimental treatment of the samples too. In the RDF of the samples, there were no multiple scattering peaks beyond ca. 6 Å which was characteristic of nanoparticle clusters of atoms. In-situ XAS, we observed differences in the samples treatment at various operating condition for pseudo radial function plot. A distinct peak at 2.0 $A^0$ possibly due to Ru-Cl scattering, this can happen at higher oxidation potential by diffusion of chloride ions inside the oxide matrix. Both ex-situ and in-situ fiting analysis we confirmed that the $RuO_2$ having rutile crystalline phase.

**Acknowledgement:-** Author SKM acknowledges DeNora Tech for full funding to carry out this research. SKM acknowledges Andrea F. Gulla for his scientific inputs and discussions. Nagappan Ramaswamy and Thomas Arruda for their support during data collection in Brookhaven National Lab. Special thanks to Brookhaven National Lab for NSLS facility to carry out ex-situ and in-situ XAS study. Special thanks to Dr. Badri Shyam for his help, discussion about XAS and continuous effort to teach SKM about data modeling. Lastly, SKM sincerely acknowledges his greatest regard to Prof Sanjeev Mukherjee for his help, guidance and scientific discussions.

**References :-**

1) T.V. Bommaraju, C.-P. Chen, and V.I. Birss, "Deactivation of Thermally Formed RuO2 + TiO2 Coatings During Chlorine Evolution: Mechanisms and Reactivation Measures," in Modern Chlor-Alkali Technology, Volume 8, edited by J. Moorhouse (Blackwell Science, Ltd., London, (2001) pp. 57.
2) S. Trasatti, Electrochimica Acta 32, 369 (1987)
3) S. Trasatti, Electrochimica Acta 45, 2377 (2000).
4) T. Helpel, F. H. Pollak and W. E. O'Grady, J Electrochem. Soc. 133, 69 (1986).
5) A.T. Kuhn and C.J. Mortimer, J. Electrochem. Soc.120, 231 (1973).
6) R.S. Yeo and J. McBreen, J. Electrochem. Soc.126, 1682 (1979).
7) R.S. Yeo, J. McBreen, A.C.C. Tseung and S. Srinivasan, J. Appl. Electrochem. 10, 393 (1980).
8) Trassati, S. et al., Electrochim Acta 1984, 29, 1503-1512.
9) Trassati, S. et al., J. Electrochem. Soc 1989, 136, 1545-1549.
10) Feng-Bin, L. et al., Electrochimica Acta 1992, 37, 2715 – 2723.
11) Subhramannia, M. et al., J. Phys. Chem. C 2007, 111, 16593-16600.
12) Devaraj, S.; Munichandriah, N.; J Phys Chem. C 2008 112 4406-4417.
13) Devaraj, S.; Munichandriah, N.; J Phys Chem. C 2008 112 4406-4417.
14) Hand Book of Chemistry and Physics 77th ed. 1996-1997 CRC press.
15) Ochoa Garcia, E.; Genesca, J. Surface Coatings and Technology 2004 184 322-330.
16) Mandelbrot, B Benoit. et.al., Nature 1984 308 721-722.
17) Feder, J.; Fractals Plenum Press, New York 1989.
18) Handbook of chemistry and Physics. 77th edition. 1996-1997 CRC press.
19) Newville, M.; J. Synchrotron Rad 2001, 8, 322-324.
20) Rehr et.al., J. Am. Chem. Soc 1991, 113, 5135-5140.
21) De Leon, M. J. et.al., Phys. Rev. B 1991, 44, 4146.
22) Fuggle, J. C.; Martensson, N.; J. Elect. Spect 1980, 21, 275.
23) Ravel, B.; Newville, M.; J. Synchrotron Rad 2005, 124, 537.
24) McKeown, A. D. et.al., J. Phys. Chem. B 1999, 103, 4825.
25) Zhan. et.al., J. Am. Chem. Soc. 2003, 125.

26) Ketchie. et.al., Chem. Mater. 2007, 19, 3406.
27) Altwasser. et.al., Microporous and Mesoporous materials. 2006, 89,109.
28) Caravati. et.al., Catalysis Today. 2007, 126, 27.

**Figures Captions:-**

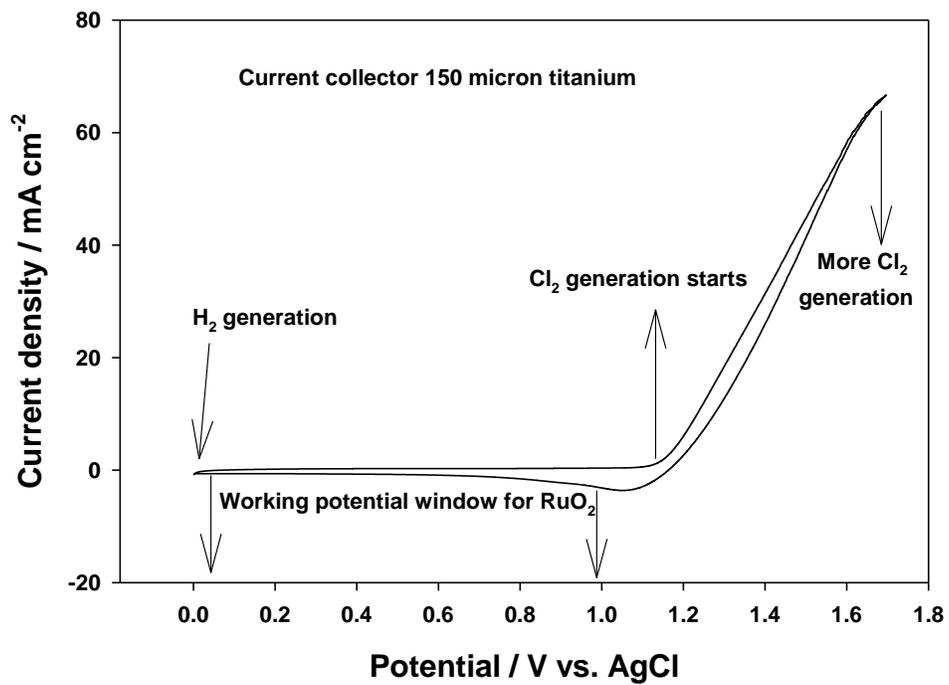

**Fig.1 Evaluation of working potential window with respect to hydrogen, oxygen and chlorine evolution reaction**

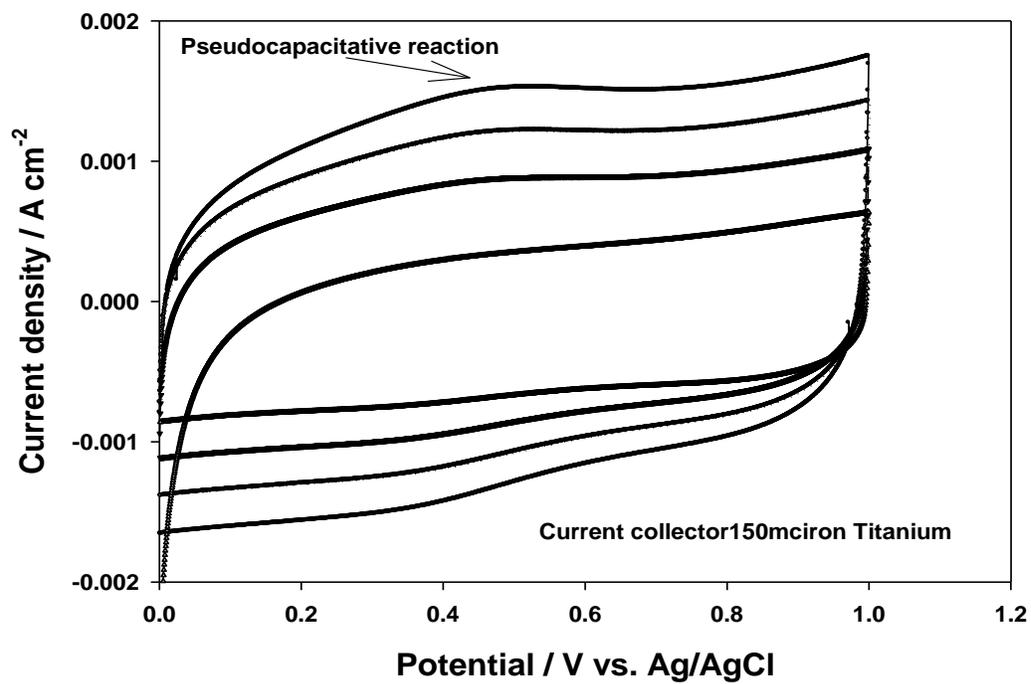

Fig.2 Cyclic voltammograms for RuO$_2$ at various scan rate.

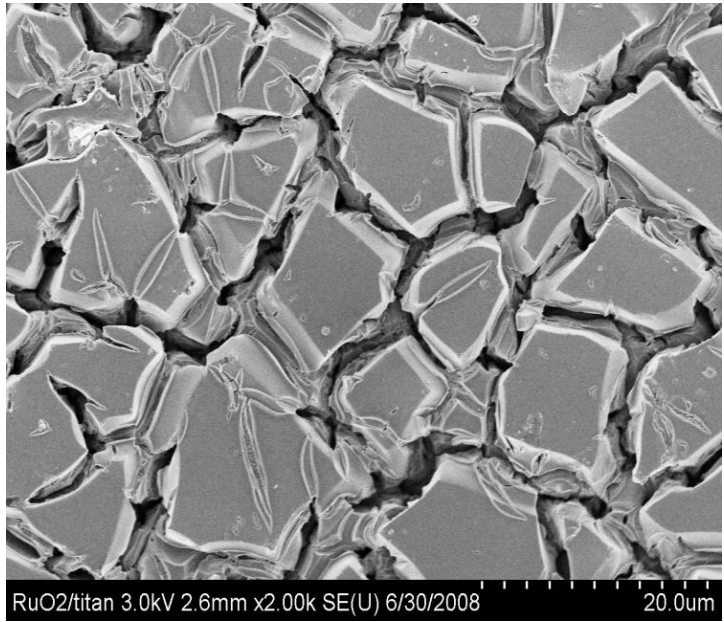

**Fig.3 SEM micrograms of RuO$_2$ coated titanium current collector.**

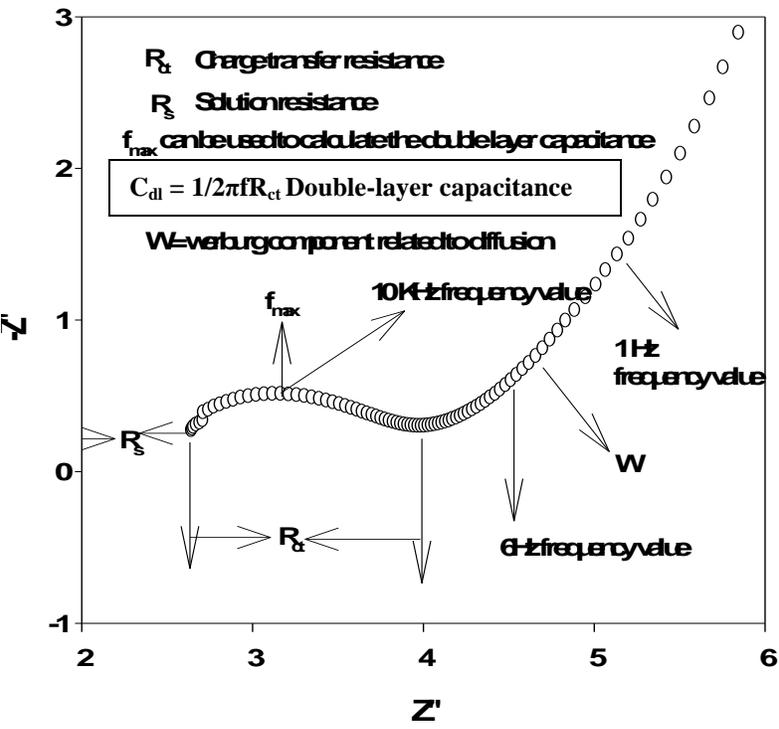
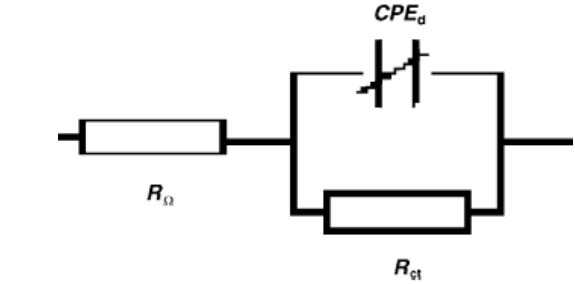

$R_\Omega$ is the solution resistance and $R_{ct}$ charge transfer resistance

**Fig. 4 (a) Model impedance diagram, Nyquist plot for Z(imp) and Z(real) and also a model for double layer. Figure 4(b) represents equivalent circuit with constant phase element. This model was used to fit the semicircle obtained in Nyquist spectrum.**

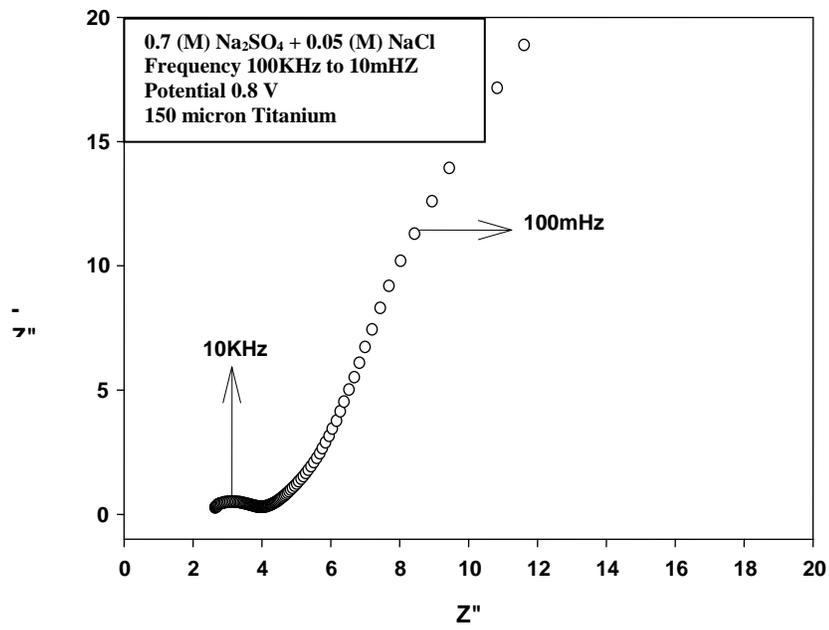

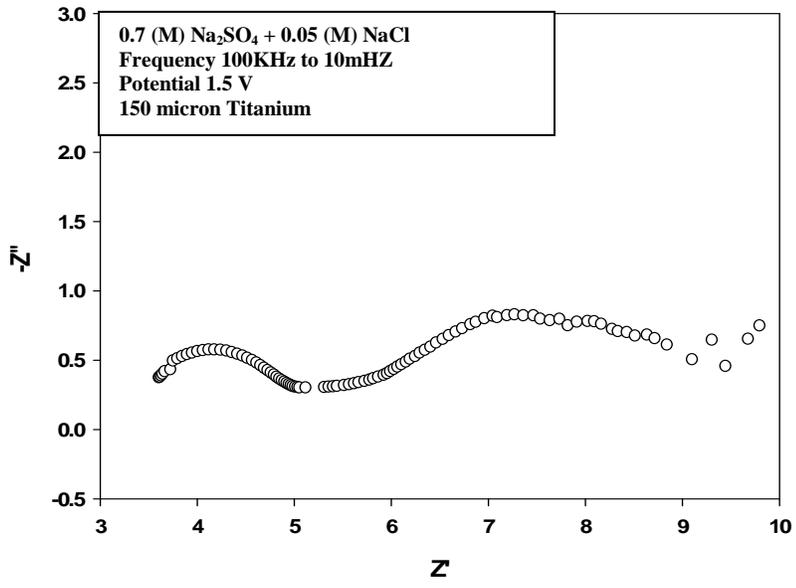

**Fig.5 Nyquist spectrum for RuO$_2$ obtained at different potential value for 150 micron thickness of titanium. Figure 5(a) & 5(b), Nyquist plot recorded at 0.8V potential and 1.5V potential.**

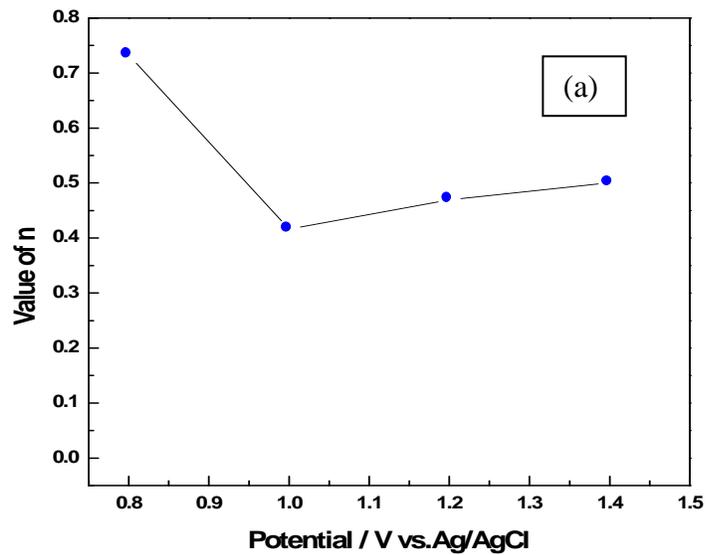

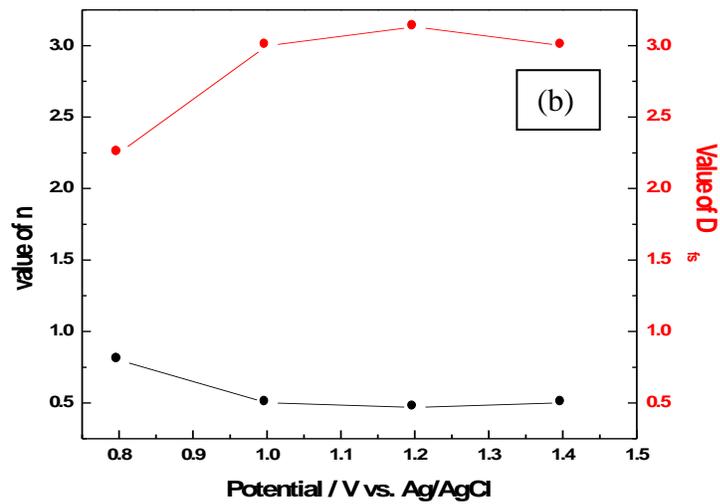

**Fig.6 (a) Plot of n value with anodic potential, n-value obtained from fitting the semi circle with equivalent circuit diagram. Fig6 (b) plot $D_{fs}$ with anodic potential. $D_{fs}$ was obtained the using n value from the fitting.**

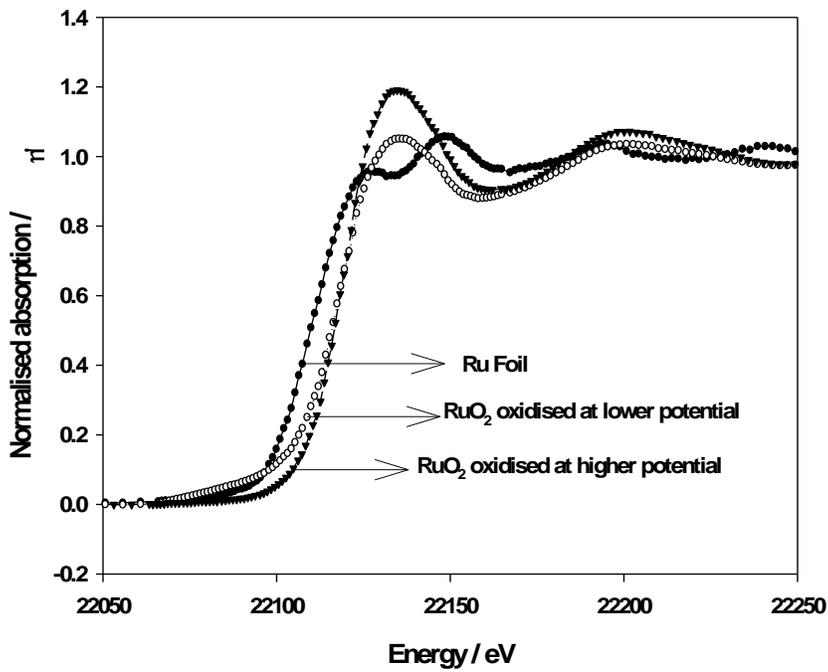

**Fig.7 XANES spectra for both samples and Ru foil.**

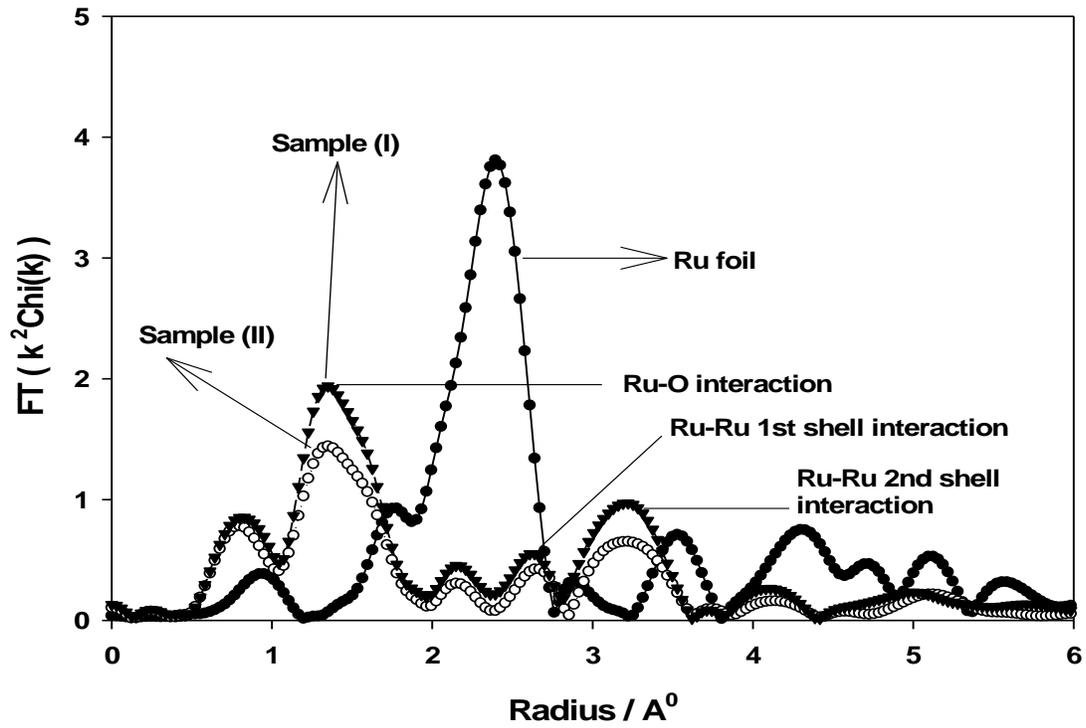

**Fig.8 Pseudo radial distribution function for both sample and Ru foil.**

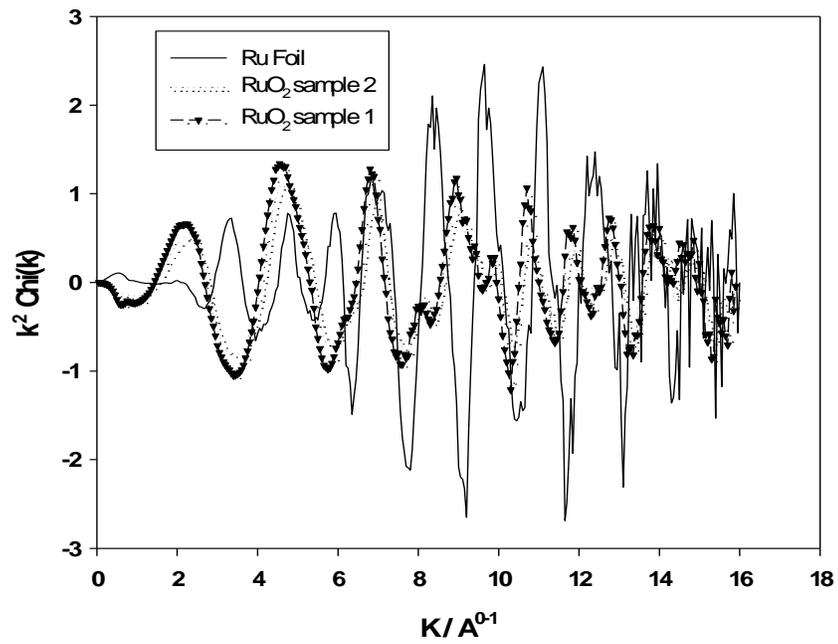

**Fig. 9 Comparison of oscillatory chi(k) function for samples and Ru foil.**

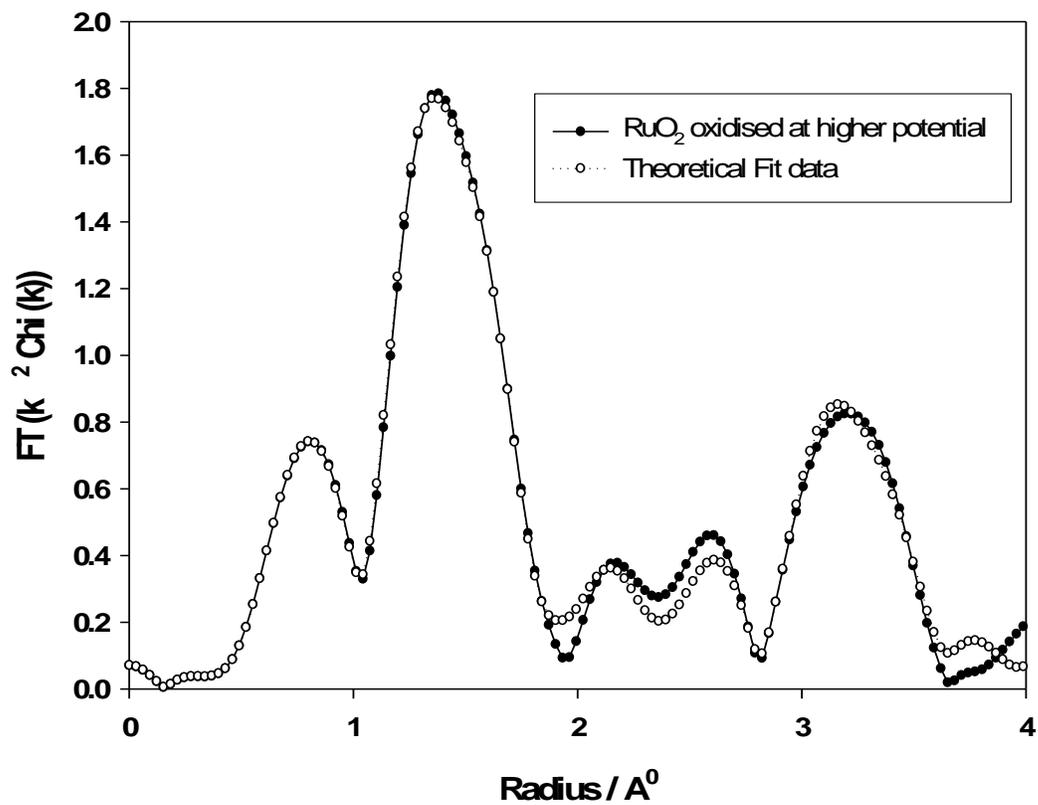

**Fig.10 Comparison of theoretical fit data with the data obtained from sample I.**

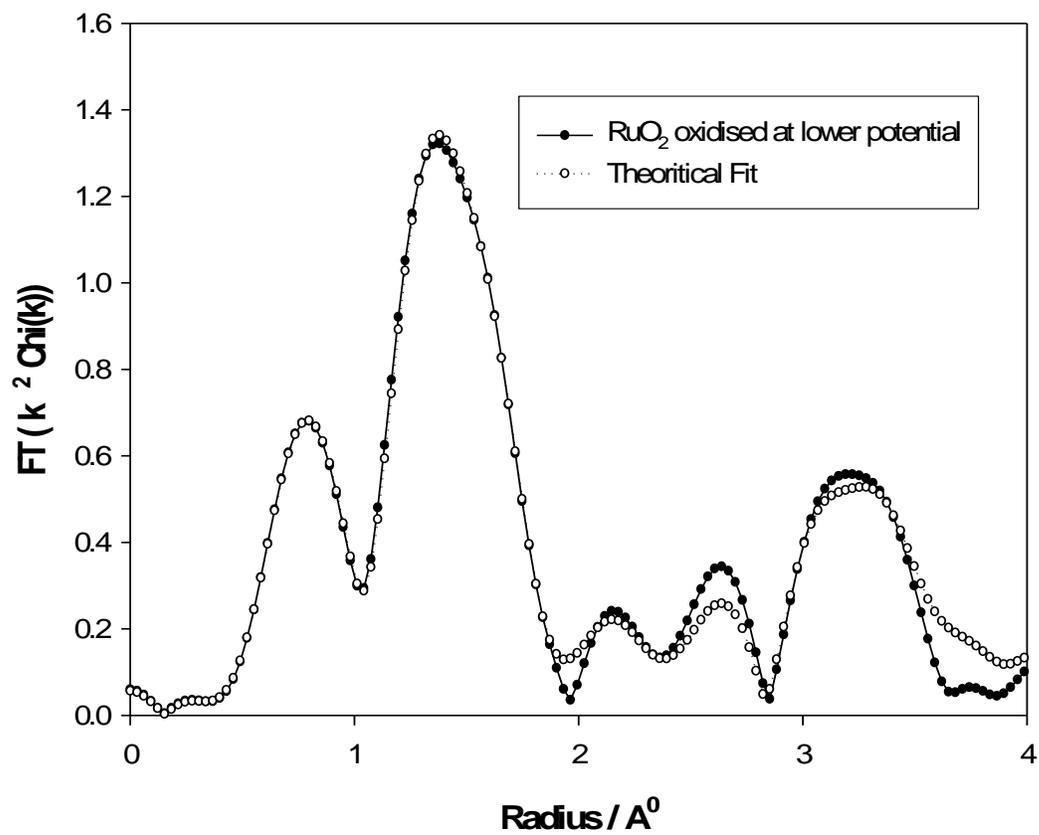

**Fig.11 Comparison of theoretical fit data with the data obtained from sample II**

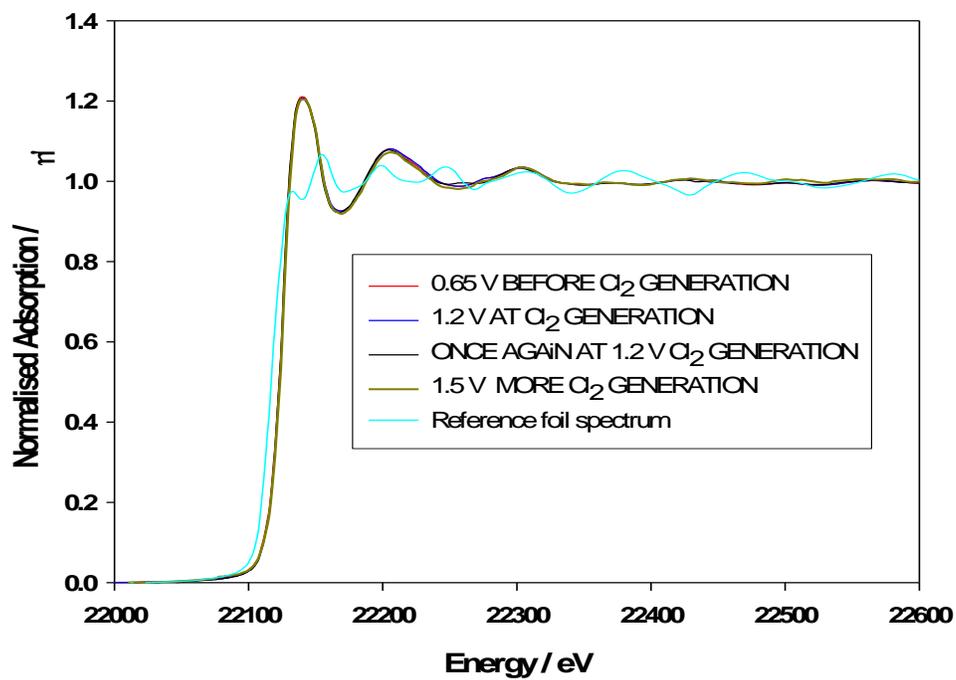

**Fig. 12** In-situ XANES results for RuO$_2$ coated over Titanium, spectra taken at different potential values.

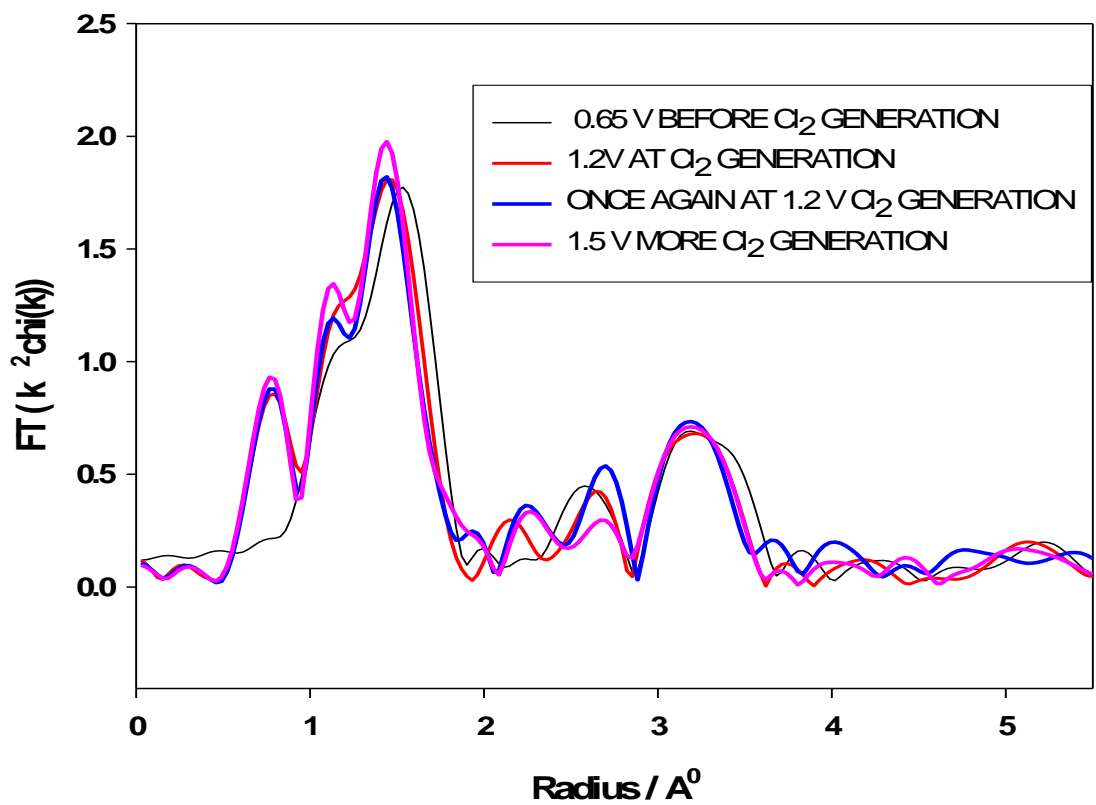

**Fig. 13 Pseudo radial function derived from in-situ experiment for $RuO_2$ sample coated over titanium.**

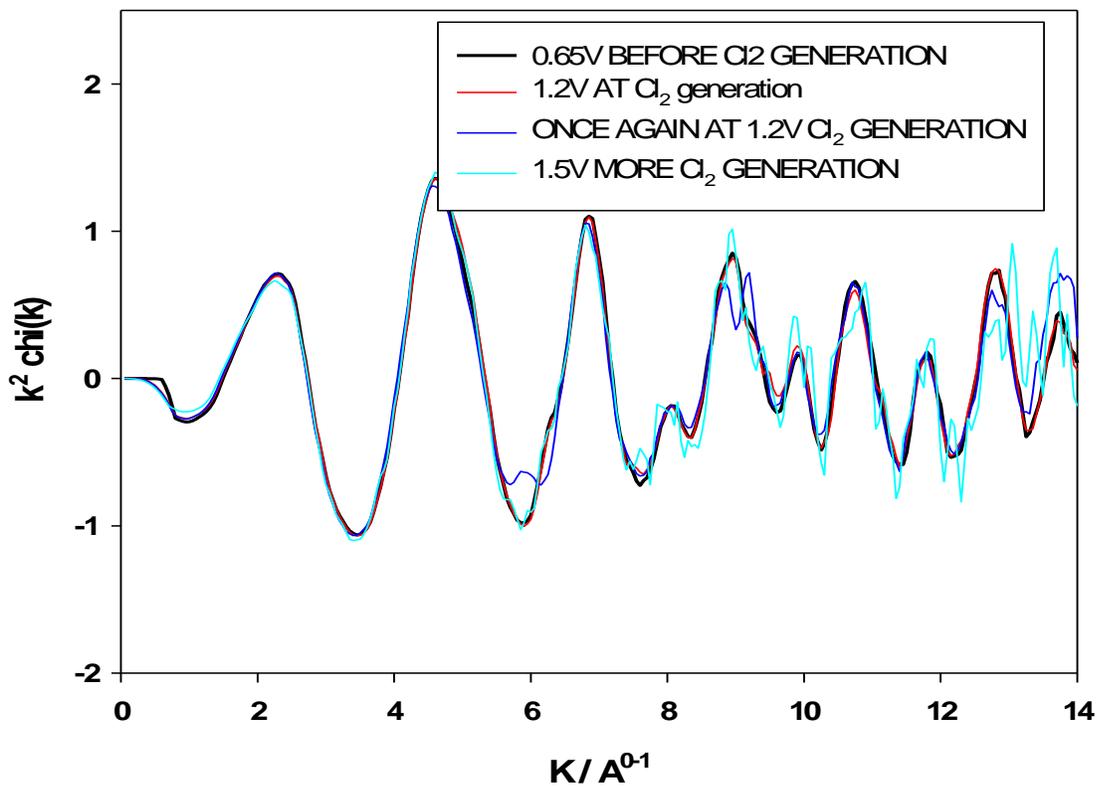

**Fig. 14 Comparison of oscillatory chi(k) function for samples obtained at various operating potential.**

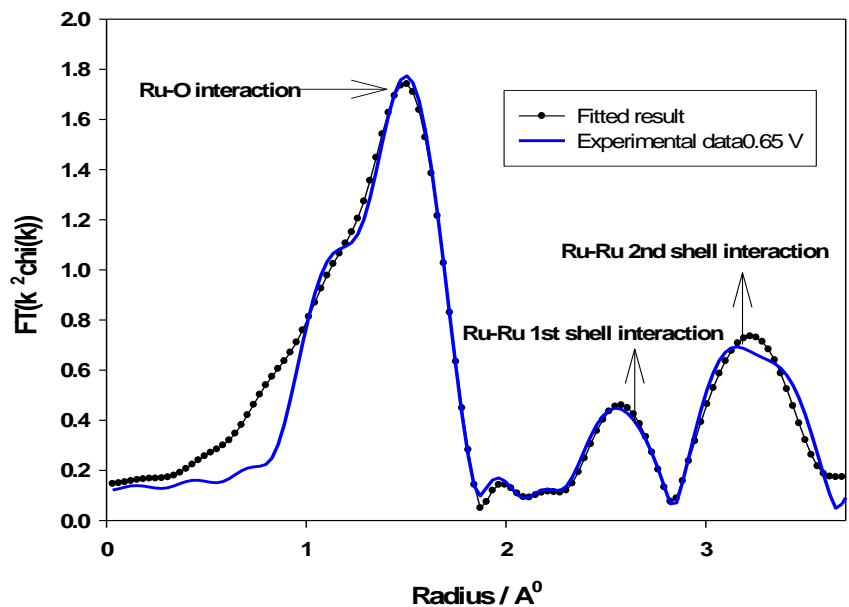

**Fig. 15 Comparison of theoretical fit and experimental results obtained at operating potential 0.65 V with respect to Ag/AgCl.**